\documentclass[a4paper,12pt]{article}
\usepackage[utf8]{inputenc}
\usepackage{cancel}
\usepackage{ulem}
\usepackage{amsfonts}
\usepackage{amssymb}
\usepackage{graphicx}
\usepackage{amsmath}
\usepackage{enumerate}
\usepackage{mathtools}
\usepackage{subfig}
\usepackage{color}
\usepackage{tikz}\usetikzlibrary{calc}
\usepackage{float}
\usepackage{here}
\usepackage{cite} 
\usepackage{mathrsfs}
\usepackage{float,epsfig}
\usepackage{dcolumn}

\usepackage{graphicx}
\usepackage{bm}
\usepackage{amsmath,amssymb,amsthm}
\usepackage[colorlinks=true,linkcolor=blue,citecolor=red]{hyperref}
\newcommand{\be}{\begin{equation}}
\newcommand{\ee}{\end{equation}}
\newcommand{\bea}{\setlength\arraycolsep{2pt} \begin{eqnarray}}
\newcommand{\eea}{\end{eqnarray}}

\def\0{{\sst{(0)}}}
\def\1{{\sst{(1)}}}
\def\2{{\sst{(2)}}}
\def\3{{\sst{(3)}}}
\def\4{{\sst{(4)}}}
\def\5{{\sst{(5)}}}
\def\6{{\sst{(6)}}}
\def\7{{\sst{(7)}}}
\def\8{{\sst{(8)}}}
\def\sst#1{{\scriptscriptstyle #1}}

\makeatletter \@addtoreset{equation}{section}

\usepackage{multirow}
\definecolor{lime}{HTML}{A6CE39}

\begin{document}

%

\title{\normalsize
{\bf 	  On   Inflationary  Models   in   $f(R,T)$   Gravity with a Kinetic Coupling  Term    }}
\author{ \small   A. Belhaj   \footnote{a-belhaj@um5r.ac.ma},  M. Benali \footnote{mohamed\_benali4@um5.ac.ma},  Y. Hassouni\footnote{y.hassouni@um5r.ac.ma}, M. Lamaaoune\footnote{mustapha\_lamaaoune@um5.ac.ma} \footnote{ Authors in alphabetical order.}
	\hspace*{-8pt} \\{\small   D\'{e}partement de Physique, Equipe des Sciences de la mati\`ere et du rayonnement, ESMaR}\\
{\small     Facult\'e des Sciences, Universit\'e Mohammed V de Rabat, Rabat,  Morocco}} \maketitle

 \maketitle
\begin{abstract}
We investigate  inflationary models in   $f(R,T)$  modified  gravity  with   a   kinetic coupling  term   $ \omega^2 G^{\mu\nu}\partial_{\mu}\phi\partial_{\nu}\phi$ having  a  positive factor needed to remove the ghosts. Taking    $f(R,T)=R+2\beta T$, we   calculate and analyse    the  relevant   observable  quantities  including the spectral index $n_s$ and  the tensor-to-scalar ratio $r$  using   the   slow-roll approximations.   Concretely, we   consider   two scenarios described by the decoupling and the coupling behaviors  between the scalar  potential  and the  $f(R,T)$  gravity via the  moduli  space by dealing with two potentials  being the quartic  one  $V(\phi) =\lambda \phi^4$ and  the small field inflation  	$V(\phi) =V_0(1- (\frac{\phi}{\mu})^\alpha)$. For the quartic  inflation model,    we  consider a decoupling behavior.   For  the small field inflation, however, we   present  the  parameter decoupling and    coupling scenarios.   For both scenarios,  we  compute and inspect  $n_s$ and  $r$  showing interesting results.  For three different values of  the number of  e-folds $N=60,65$ and $70$, we   find that   the  coupling  between $f(R,T)$ and the scalar potential  via the  moduli space  provides  an excellent agreement with the  observational findings.   In the last part of this work, we provide   a possible  discussion on   the amplitude
of  scalar power spectrum needed  to  provide  a  viability of the proposed  theory.  Considering   the second potential form in the parameter coupling scenario, we find  acceptable values  in certain points of the moduli space.\\

{\bf Keywords}: Inflation,  Kinetic coupling terms,   Cosmological observables, Slow-roll computations, Observational data.

	\end{abstract}


\newpage
\section{Introduction}
Inflationary models  have been extensively  investigated   by considering various  gravity   theories  including the modified  ones \cite{r1,r2,r4,r26,r27}.  Certain elaborated ones have brought  roads   to understand the universe evolution   features such as    horizon, flatness and large structure issues\cite{r6,r7,r8}.  Several theories have been proposed by considering   the  single and  the multiple scalar fields via  the  potentials in order to     to discuss  standard cosmological problems within  general relativity (GR).  The simplet ones involve  only a single field being interpreted  as a dominated element to drive inflation.  Such an object has been approached by dealing   with  various theories  even the ones appearing in higher dimensions  \cite{r9,r11,r112,r12}.\\
Developments in string theory  has been also exploited to engineer  inflationary  models    relying on the D3-brane physics  using  the   Randall-Sundrum II (RS-2) mechanism.   In this way,   interesting  scalar  potentials  involving certain  parameters such as   the brane tension have been studied where the stringy corrections of the involved quantities have been derived \cite{r13,r14,r15,M1,M2}.   In particular, the compactifications of superstring models and M-theory  have  been explored producing  many scalar fields. These objects  are associated with geometric deformations of the metric and other non trivial tensor fields coupled to  the brane objects\cite{r16,r17,r19}.   The   corresponding cosmological quantities have been approached and examined. These   scalar fields have been exploited  to confront  the  theoretical predictions with  observational data provided by cosmic microware backgrounds (CMB) and  the Planck experimental  results \cite{r20,r21,r22}.\\
Alternatively,   GR  modified  models have been explored  showing interesting  inflationary results.  The most dealt with   ones are    $f(R)$  modified gravity theories   where $R$  denotes  the Ricci scalar  being largely  studied  by taking several  scalar potentials \cite{r23,r24,r25}.   Based on   the  slow-roll  analysis, the  associated   spectral index $n_s$  and  the tensor-to-scalar ratio  $r$ have been  calculated   with specific values of   the  number  of e-folds  needed to bridge the resulting  models  with  the  observational findings.   These models have been extended by adding others  quantities including  the trace $T$ of  the stress-energy tensor.  The resulting   is called   $f(R,T)$  gravity theory  being   extensively   studied   providing certain  inflationary models \cite{nsr,nsr1}. 
 A close examination reveals that  the modified gravity theories  are motivated by  the study of  dark energy (DE).  It has been remarked that the latter   could  be exploited to bring  certain explanations of the accelerating   expansion aspect  of the universe.  Concretely,   the  modified gravity theories could  generate   certain dynamical  behaviors  which  can be interpreted as  DE contributions \cite{d1,d2,d3}.  These behaviors  go  beyond the cosmological constant in the GR context.\\
Alternatively,   a  kinetic coupling  term   has been  proposed   to generate   inflationary models motivated by Higgs field physics.  In particular,  the relevant  observable   quantities have been computed  and analyzed  \cite{r28, r29,r30,r31,ref1,ref2,ref3}.  In the absence parameter coupling scenarios,  however, certain models do not provide observational features.

It has been observed that the form of the scalar potential  can play  a primordial  role in  the building  of 
inflationary models   derived from   various  theories including   superstring  models, and M-theory. The choice of the  potential form usually  depends on    motivations relaying on  known  models such  as  the  standard model (SM) of the  particle physics \cite{ref1,ref2}.      It  has been remarked  that famous illustrations are  the 
chaotic inflation potential  and the minimal supersymmetric standard model (MSSM) inflation potential.
In addition to these studied models, other forms of the scalar potential have been  dealt with  in connections with 
 black hole physics and quintessential  DE scenarios \cite{d4,d5,d6}.

The  main objective of this work is to contribute to these activities by  studying   inflationary models   within   the  $f(R,T)$  modified gravity  by   means of a    kinetic coupling  mechanism   via  the term   $ \omega^2 G^{\mu\nu}\partial_{\mu}\phi\partial_{\nu}\phi$,   with a positive factor  needed to remove the ghosts.   Taking $f(R,T)=R+2\beta T$,  we reconsider the study of   the  quartic  potential  and the  small field  one    given by $V(\phi) =\lambda \phi^4$ and   	$V(\phi) =V_0(1- (\frac{\phi}{\mu})^\alpha)$, respectively.     Within such a  mechanism, we  compute and examine     the  relevant  observables    such as  the spectral index $n_s$ and  the tensor-to-scalar ratio $r$  by the help of    the  slow-roll approximations.  Precisely, we provide two situations described by the decoupling and the coupling between the scalar  potential  and the modified  gravity via the  involved moduli   space.  In the first model associated with   the quartic potential,  we  investigate   a    decoupling behavior.  In the second one  corresponding to   the small field inflation, however, we study  the  parameter decoupling and    coupling    scenarios.   In  both  scenarios,  we investigate   the spectral index $n_s$ and the tensor-to-scalar ratio $r$.   Precisely, we illustrate   the associated cosmological behaviors  using graphic presentations.  For three different values of the e-folding    number $ N = 60, 65$  and 70, we find that the coupling between  the $f(R,T)$ modified   gravity and the scalar  potential  via the moduli space  provides  an excellent   agreement with observational findings.    We end this work by elaborating    a possible  discussion on   the amplitude
of  scalar power spectrum  which could be needed  to  provide  a  viability of the proposed  theory. \\ 
The organization of this paper is as follows. 
In section 2, we  study     $f(R,T)=R+2\beta T$   gravity action  within a  kinetic  coupling  scenario  and  establish  the  field  equations of motion. In section 3,  we investigate the  relevant inflation parameters in such a modified gravity model with a generic scalar potential. In section 4,  we make contact with observational data by considering two models  being the small field  and the  quartic potential inflations  using the  parameter decoupling and   coupling scenarios.  In section 5, we  give  concluding remarks.  In this  work, we take  $G=c=1$ corresponding to the  
gravitational constant and the light  speed,  respectively.
\section{Field equations  in $f(R,T)$ gravity  with a  kinetic coupling  term }
 In this section, we   give   a concise   discussion  on  $f(R,T)$ gravity   with a  kinetic coupling  term. 
Then, we  present the associated  field equations needed to investigate  inflation models in the resulting theory.  To start, we consider   the following  action 
\begin{equation}
S = \int dx^4\sqrt{-g}\left( \frac{f(R,T)}{16\pi}+{\cal L}_m\right)
\end{equation} 
where $ f(R,T)$ is an arbitrary function of  the Ricci scalar  $R$ and  $T$ denoting   the trace  of  the stress-energy tensor  $T_{\mu\nu}$  of the matter sector   and where  $g$  is the determinant of the metric $g_{\mu\nu}$ \cite{T1,T2,T3,T4,T5,T6}. In the absence of   the kinetic coupling  terms,    ${\cal L}_m$  describes  a  matter Lagrangian density given by 
\begin{equation}
{\cal L}_m= -\frac{1}{2}g^{\mu\nu} \triangledown_{\mu}\phi\triangledown_{\nu}\phi - V(\phi)
\end{equation} 
where $\phi$ indicates   the dynamical  scalar field    controlled  by  a potential $V(\phi)$.  In this situation, the stress-energy tensor  $T_ {\mu\nu}$  of the matter reads as 
\begin{equation}
T_{\mu\nu}=-\frac{2}{\sqrt{-g}} \frac{\delta(\sqrt{-g}{\cal L}_m)}{\delta g^{\mu\nu} }.
\end{equation} 
In this work,  we combine two scenarios to provide inflationary models.   The first  scenario that we   follow   here is based on the  introduction  of   a  kinetic coupling  term in  the previous action. This term has been extensively  studied   in connections   with 
Higgs theory  coupled to to the  gravity which could  appear  in heterotic  superstring models and related topics \cite{ST1}.  In this way, we consider  the matter sector described by  the following  Lagrangian density 
\begin{equation}
{\cal L}^k_m= -\frac{1}{2}(g^{\mu\nu} - \omega^2 G^{\mu\nu})\triangledown_{\mu}\phi\triangledown_{\nu}\phi - V(\phi)\end{equation} 
where $G_{\mu\nu}$  is  the Einstein tensor and  $\omega$ is a parameter  having  the dimension of  the mass scale.  In present analysis, we  take a positive sign in  the action corresponding to   the factor  $+ \omega^2 $.  The positive sign of the kinetic coupling term ensures the absence of the ghost in the underlying  models \cite{ref1,ref2}.   A close inspection shows that  the kinetic terms have  been  introduced  in order to reduce  the tensor-to-scalar ratio $r$ providing certain arguments with CMB observations.
Varying the modified  action with respect to the metric  $g_{\mu\nu}$,  we can obtain  the following field equation
\begin{eqnarray}
	\label{Eq}
	f_R (R,T)R_{\mu\nu}&-&\frac{1}{2}g_{\mu\nu} f(R,T)-(g_{\mu\nu} \square-\triangledown_\mu\triangledown_\nu)f_R(R,T)\\\nonumber
	&=&-f_T(R,T)T_{\mu\nu} - f_T(R,T)\Theta_{\mu\nu} +8\pi(T^{(\phi)}_{\mu\nu} +\omega^2 A_{\mu\nu}  )
\end{eqnarray}
where  we have used the notion   $f_T(R,T)= \frac{\partial  f(R,T)}{\partial T}$ 	 and $f_R(R,T)= \frac{\partial  f(R,T)}{\partial R}$. The  involved quantities are found to be    
\begin{eqnarray}
\Theta_{\mu\nu} &= & g^{\alpha\beta}\frac{\delta T_{\alpha\beta}}{\delta g^{\mu\nu}}\\
T^{(\phi)}_{\mu\nu}&= &\triangledown_\mu\phi\triangledown_\nu\phi-\frac{1}{2}g_{\mu\nu}\triangledown_\rho\phi\triangledown^\rho\phi+g_{\mu\nu}V(\phi)\\		
A_{\mu\nu}&=&-\frac{1}{2}\triangledown_\mu\phi\triangledown_\nu\phi R + 2\triangledown_\alpha\phi\triangledown(_\mu\phi R^\alpha_\nu)+\triangledown^\alpha\phi\triangledown^\beta\phi R_{\mu\alpha\nu\beta}\nonumber\\&+&\triangledown_\mu\triangledown^\alpha\phi\triangledown_\nu\triangledown_\alpha\phi - \triangledown_\mu\triangledown_\nu\phi\square\phi - \frac{1}{2}(\triangledown\phi)^2 G_{\mu\nu}\\&+& g_{\mu\nu}[-\frac{1}{2}\triangledown^\alpha\phi\triangledown^\beta\phi\triangledown_\alpha\phi\triangledown_\beta\phi+\frac{1}{2}(\square\phi)^2-\triangledown_\alpha\phi\triangledown_\beta\phi R^{\alpha\beta}].\nonumber
\end{eqnarray}
Contracting Eq.(\ref{Eq}), the Ricci scalar $R$ and the trace $T$ of the stress-energy tensor are linked via the relation 
\begin{eqnarray}
	f_R (R,T)R - \frac{1}{2} f(R,T)=-f_T(R,T)T - f_T(R,T)\Theta +8\pi(T^{(\phi)} +\omega^2 A  )
\end{eqnarray}
where one has used  $A = A^\mu_\mu$ and $\Theta=\Theta^\mu_\mu$.  In this framework, it has been remarked that   the covariant derivative of the energy-momentum tensor is not null  $\triangledown_\mu T_{\mu\nu}\neq 0$. In order to obtain the modified continuity relation,  the covariant derivative of   (\ref{Eq}) should be  performed. Indeed, one writes 
 \begin{eqnarray}
 	\label{CD}
 \triangledown^\mu \bigg[ f_R (R,T)R_{\mu\nu}&-&\frac{1}{2}g_{\mu\nu} f(R,T)-(g_{\mu\nu} \square-\triangledown_\mu\triangledown_\nu)f_R(R,T)\\\nonumber
 	&=&-f_T(R,T)T_{\mu\nu} - f_T(R,T)\Theta_{\mu\nu} +8\pi(T^{(\phi)}_{\mu\nu} +\omega^2 A_{\mu\nu}  )\bigg] .
 \end{eqnarray}
Using the identities $\triangledown R_{\mu\nu} - \frac{1}{2}Rg_{\mu\nu} = 0$, and $(\triangledown_\nu\square-\square \triangledown_\nu) f_R(R, T ) = R_{\mu\nu}\triangledown^\mu f_R$ and  considering  the following relation 
 \begin{eqnarray}
	\triangledown^\mu \bigg[ f_R (R,T)R_{\mu\nu}&-&\frac{1}{2}g_{\mu\nu} f(R,T)-(g_{\mu\nu} \square-\triangledown_\mu\triangledown_\nu)f_R(R,T)\bigg] =0
\end{eqnarray}
  the divergence of the stress-energy tensor $ T_{\mu\nu}$  takes the form 
\begin{eqnarray}
\triangledown^\mu T_{\mu\nu} = - \frac{1}{f_T(R,T)}\left[   (T_{\mu\nu}+\Theta_{\mu\nu}) \triangledown^\mu f_T(R,T) - 8\pi(\triangledown^\mu T^{(\phi)}_{\mu\nu} +\omega^2 \triangledown^\mu A_{\mu\nu}) \right].  
\end{eqnarray}
For  $\omega=0$,  we recover the usual  expressions \cite{T1,T2}.  Exploiting the pressure $p$, we consider  the following situation 
\begin{equation}
		\Theta_{\mu\nu} = -2T_{\mu\nu}-p g_{\mu\nu}
\end{equation}
associated with  a   homogeneous and an isotropic  universe which can be modeled  by  a perfect fluid \cite{T1,T2,T3}. 
 
A close examination reveals  that  a generic form of    $ f(R,T)$  generate complex computations leading to a very hard task.  However, certain models have been proposed by dealing with  special functions recovering the GR  limit  where  $ f(R,T)$ reduces to $R$. Ignoring  the  $RT $ mixing terms,  such models are based on the following forms 
\begin{equation}
\label{frt}
	f(R,T)= R\,+2g(T)
\end{equation} 
where  $g(T)$  is an arbitrary function of $T$. 
Motivated by DE activities,  the second scenario relay on  a  possible extension of GR by restricting  the present investigation to a linear form of    $ f(R,T)$ gravity  given by 
\begin{equation}
\label{beta}
	f(R,T)= R\,+\,2\,\beta T
\end{equation} 
where  $\beta$ is a new positive parameter  being  independent of  $T$ and  $R$. It can be  considered as   a relevant  modified  gravity parameter. As advantages, this form can be considered as an economic extension of GR without  RT mixing terms  which  need more complicated  calculations.  Moreover, it involves only one extra parameter reducing the generic moduli space. As envisaged, this parameter will be exploited to suggest a new coupling between the gravity and the dynamical scalar field through   the potential $V(\phi)$.  This coupling parameter mechanism being elaborated  by the implementation of  $\beta$ in  $V(\phi)$  could open  possible roads    to establish bridges with DE modeling  from inflation investigations. 
Using  $g_{\mu\nu}$ and $\phi$ variations  via  the Friedman-Lemaitre-Robertson-Walker (FLRW) metric, we get the equations of motion
 \begin{eqnarray}
3 H^2 &=&4\pi\dot{\phi}^2(1 + 9\omega^2 H^2) + (8\pi+6\beta) V(\phi)-4\beta\dot{\phi^2}\\
2\dot{H} + 3 H^2 &=& - 4 \pi\dot{\phi}^2 (1 - \omega^2(2\dot{H}+3H^2 + 4H\ddot{\phi}\dot{\phi}^{-1})) + (8\pi +6\beta) V(\phi)-4\beta\dot{\phi}^2\\
(\ddot{\phi} +3H\dot{\phi}) &+& 3\omega^2( H^2\ddot{\phi} + 3H^3\dot{\phi} +2 H\dot{H}\dot{\phi}) = - V'(\phi)
\end{eqnarray}
where the  dot is the derivative with respect to  the cosmic time $t$,  and one has used $V'=\frac{dV}{d\phi}$.  $H$ is the Hubble parameter defined by $H=\frac{\dot{a}}{a}$ where  $a(t)$ is  a scalar factor.  For $\beta=0$, we recover the equations  of motion  in the RG limit described by   $f(R,T)=R$   \cite{r28,r31}. A close examination shows  that the cosmological  observables can be obtained using certain approximations. To make contact with the observational data \cite{r20,r21,r22},   the  appropriate  approximations will be  exploited. 
\section{ Slow-roll computations}
 To get  the observable quantities in an acceptable way,  we  could  use  certain  approximations simplifying the previous equations of motion by removing the most slowing changing contributions \cite{SR}. Under  the  slow-roll  approximations which could provide  exact solutions of the   equations  of motion,  the source of the energy is the scalar potential being  the dominated pieces in the inflation scenario. Moreover, the acceleration  scalar field  terms  could be neglected  in such equations of motion even  in generic  situations  where the scalar  potential is not specified. 
 Concretely,  we calculate the relevant inflation  parameters  associated with the  above  action involving the proposed  $f(R,T)$ gravity function   in  the kinetic term mechanism. 
In   the  inflation phase, the slow-roll parameters are  given by 
\begin{equation}
	\epsilon = - \frac{\dot{H}}{H^2}, \quad
	\eta = \frac{\dot{\epsilon}}{H \epsilon}, \quad
	\kappa_{0}  = 12 \pi \kappa \dot{\phi}^2, \quad 
	\kappa_{1}  = \frac{\dot{\kappa_{0}}}{H \kappa_{0}}.
\end{equation}
According to the slow-roll inflation mechanism, these parameters are constrained  as follows
\begin{equation}
	\epsilon, \eta, \kappa_{0},\kappa_{1} <<1.
\end{equation}
In such  slow-roll approximations, the field equations of motion  reduce to 
\begin{align}
3H\dot{\phi}   =&-9\omega^2 H^3\dot{\phi}-V'(\phi)\\
 3H^2 = & (8\pi +\,6\,\beta)V(\phi)\\
\dot{H} =& - 4 \pi\dot{\phi}^2 (1 +3\omega^2 H^2).
\end{align}
 Calculations reveal that  these field  equations can be solved  as follows 
\begin{align}
	\label{a3}
	H^2 &=(2\beta +\frac{8 \pi }{3}) V(\phi )\\
		\label{a2}
	\dot{H}&= -\frac{2 \pi  V'(\phi )^2}{3 (6 \beta +8 \pi ) V(\phi ) \left(1+(6 \beta +8 \pi )  \omega ^2 V(\phi )\right)}	\\
\label{a1}
\dot{\phi} &=-\frac{V'(\phi )}{\sqrt{6(9 \beta +24 \pi )} \sqrt{V(\phi )} \left(1+(6 \beta +8 \pi ) \omega ^2 V(\phi )\right)}.
\end{align}
 Exploiting the   slow-roll parameters,  Eq.(\ref{a1}), Eq.(\ref{a2}), and Eq.(\ref{a3}), we can get the relevant inflation parameters  as functions of  $V(\phi)$,  $\beta$  and  $\omega$. Indeed,   we obtain 
\begin{align}
\epsilon &=\frac{\pi  V'(\phi )^2}{(3 \beta +4 \pi )^2 V(\phi )^2 \left(1+ (6 \beta +8 \pi ) \omega ^2 V(\phi )\right)} \\
\eta &=\frac{V'(\phi )^2 \left(1+ (9 \beta +12 \pi ) \omega ^2 V(\phi )\right)-V(\phi ) V''(\phi ) \left(1+ (6 \beta +8 \pi ) \omega ^2 V(\phi )\right)}{(3 \beta +4 \pi ) V(\phi )^2 \left(1+(6 \beta + 8 \pi ) \omega ^2 V(\phi )\right)^2}.
\end{align}
In the  inflationary  building models,  the  slow-roll parameters  provide  the scalar field value at the end of the expansion $\phi_E$ by imposing the constraint $\epsilon(\phi_E)=1$. Moreover,    we can find the beginning of inflation from  the total logarithmic phase.  It is convient to express  the  above  slow-roll parameters   in terms of  the e-folding number corresponding to   the duration of inflation.  This number can be  computed via the following relation 
\begin{eqnarray}
N = \int_{t_I}^{t_E}{ H dt } =  \int_{\phi_I}^{\phi_E}{ \frac{H }{\dot{\phi}} d\phi } 
\end{eqnarray}
where the subscript $I$ and $E$ denote the  associated   values at the onset and the  offset time of inflation, respectively. The calculation gives 
\begin{eqnarray}
N= (6\beta +8 \pi )\int_{\phi_I}^{\phi_E}\frac{V(\phi) \left(1+(6 \beta +8 \pi ) \omega ^2 V(\phi)\right)}{V'(\phi)}d\phi.
	\end{eqnarray}
 In order to solve the  big-bang issues, it is usually required  to  consider  $N \geq  60$. \\
Roughly,  the main goal  of the  gravitational theories is to provide consistent predictions  matching  with  the observational  data.  It has been remarked that  the implementation  of   $f(R,T)$ in   the action does  not effect the primordial cosmological quantities  $n_s$ and $r$ \cite{nsr,nsr1}.  However,   the non-minimal kinetic term coupling  can affect  such quantities.  According to \cite{r29,r30}, the scalar spectral index $n_s$ and the  tensor-to-scalar ratio $r$ are expressed as follows
\begin{eqnarray}
	n_{s} -1& =& -2\epsilon - \eta\\
		r &=& 16 \epsilon.
\end {eqnarray}
In the present  modified gravity model, these quantities  are found to be
\begin{eqnarray}
\label{a5}
n_{s}  &=&1-\frac{2 \pi  V'(\phi )^2}{(3 \beta +4 \pi )^2 V(\phi )^2 \left(1+ (6 \beta +8 \pi ) \omega ^2 V(\phi )\right)}\\\nonumber&+&\frac{V(\phi ) V''(\phi ) \left(1+(6 \beta +8\pi ) \omega ^2 V(\phi )\right)-V'(\phi )^2 \left(1+3 (3 \beta +4 \pi ) \omega ^2 V(\phi )\right)}{(3 \beta +4 \pi ) V(\phi )^2 \left(1+(6 \beta +8 \pi ) \omega ^2 V(\phi )\right)^2}\\
\label{a6}
r &=&\frac{16 \pi  V'(\phi )^2}{(3 \beta +4 \pi )^2 V(\phi )^2 \left(1+ (6 \beta +8 \pi ) \omega ^2 V(\phi )\right)}.
\end{eqnarray}
Having calculated the primordial observables,  we move to  study    certain     models  specified by the  potential form. To  test such models, we   provide  bridges  to   observational findings and results.
\section{Inflation   modeling in  $f(R,T)$ gravity  with    a kinetic coupling  scenario}
 In  this section, we would like  to  investigate concrete  inflation models  with a  kinetic   coupling term in  the $f(R,T)$  modified gravity.   It has been remarked that the previous computations depend  on   $\beta$  and  the  potential  function   of  the single scalar field.  However, interesting ones should  provide  smallness numerical values of the involved cosmological  quantities.  In the present investigations, the  proposed modified   gravity will be  controlled  by varying  $\beta$ in the   scalar potential  $V(\phi)$.   This idea could   provide inflationary    models going beyond the previous ones   generating  interesting  results  compared   with models where the gravity parameters  and the scalar  potential  ones are not  related.   In the   parameter   coupling mechanism  of  the modified   gravity with  the matter sector via the potential, the   cosmological observables  are  modified  in terms of $\beta$.   Instead of being general, we deal with     two  different potentials and make contact with observational data.  The choice  such  potentials   could be motivated by  certain parallel  investigations  including  either the coupling of the  standard  models with the gravity  or DE activities.    It has been observed that the  possible choices could   not only  rely on the known  scalar field potentials but also on the corresponding    possible contact with observational constraints   obtained  in different  inflation scenarios. 
Motivated by the potentials  explored in  the study of Higgs  inflation  with a kinetic term \cite{r112,V1}, we  investigate two models  known by the  quartic  potential and the  potential associated with the  small field inflation. Then, we provide a comparative discussion.
\subsection{Quartic potential in   kinetic   coupling scenarios} 
\begin{figure}[!ht]
	\centering
	\begin{tabbing}
		
		\hspace{5.6cm}\= \hspace{5.6cm}\=\kill
		
		\includegraphics[scale=0.60]{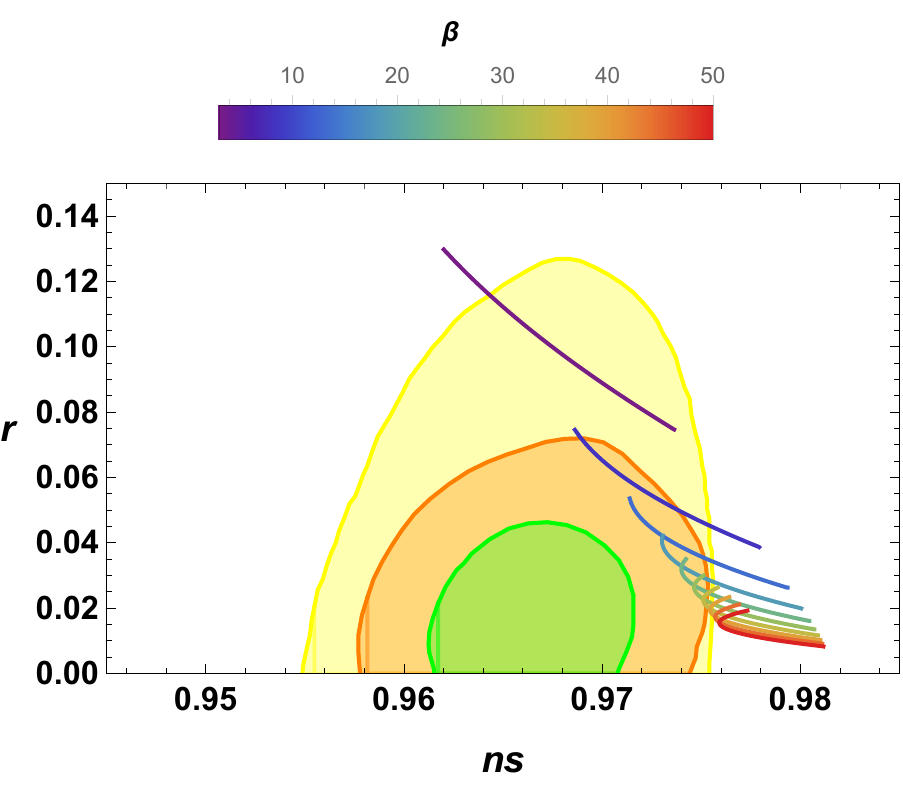} \>
		
		\includegraphics[scale=0.60]{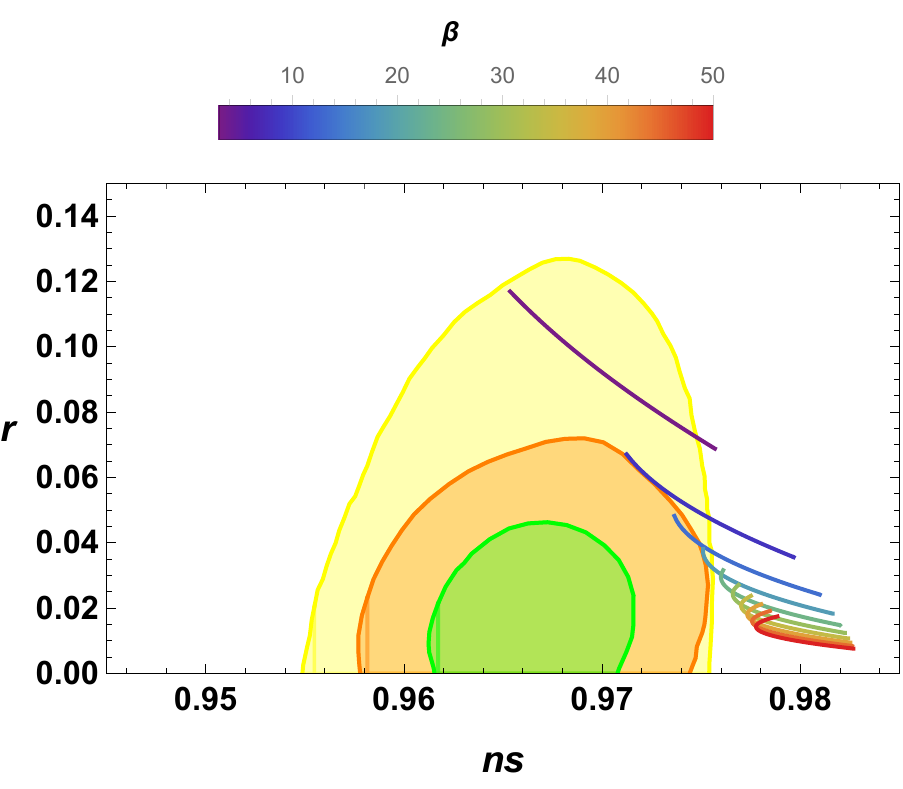} \>
		
		\includegraphics[scale=0.60]{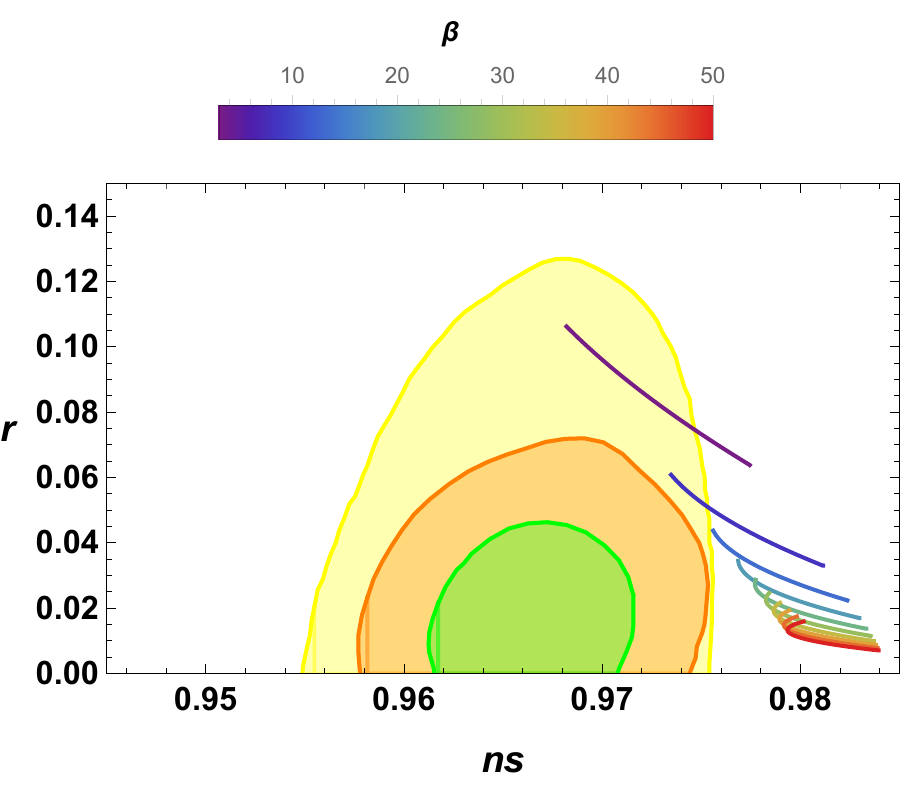} \\
		
	\end{tabbing}
	
	\vspace{-1.5 cm}
	
	\caption{\it \footnotesize  The behavior of one dimensional  curves $n_s-r$ by varying the Parameter $\beta=1,\ldots, 50$ and the parameter $x=0.001,\ldots, 1$. The plots from left to right correspond to the number e-folding  $N=60$,  $N=65$ and  $N=70$, respectively.  The green and the orange contour constraints  represent  68\% and 95\% confidential levels of Planck results (TT,TE,EE+lowE+lensing+BK15+BAO), respectively. The yellow contour is associated with  the Planck results (TT,TE,EE+lowE+lensing).}
	\label{f1}
\end{figure}

The first  model  that we would like to  deal with  is the   inflation potential  investigated in \cite{V1}. This concerns the quartic potential  given by 
\begin{equation}
	\label{V2}
	V(\phi) = \lambda  {\,\phi^4}
\end{equation}
where $\lambda$ is  a  dimensionless  coupling  parameter.  This parameter can be interpreted as the scalar self-coupling. In connection with SM, this potential has been exploited to make contact with the Higgs field considered as the only scalar field in the underlying physics.  In this way, the dynamical  scalar   $\phi$ could  be identified  with the Higgs field. It is worth noting that other scenarios could be possible via intermediate interactions.  To get  the scalar spectral index $n_s$ and the tensor-to-scalar ratio $r$ as functions of $V_0$ and $\phi$,   we should  compute  the  e-folding number. For this model,  it takes the following form 
\begin{equation}
N = \frac{1}{24} B \left(\phi_I^2 -\phi_E^2\right) \left(3\,+\,B x \left(\phi_E^4+\phi_E^2 \phi_I^2+\phi_I^4\right)\right)
\end{equation}
where one has used	$B = 8 \pi + 6 \beta $  and    $x=\lambda \,\omega^2$. \\
The scalar spectral index $n_s$ and  the tensor-to-scalar ratio $r$  are  found to be 
\begin{eqnarray}
r &=&\frac{1024 \pi }{B^2 \phi ^2 \left(1\,+\,B\, x\, \phi ^4\right)}\\
n_s&=&1-\frac{128 \pi }{B^2 \phi ^2 \left(1 + B x \phi ^4\right)}-\frac{8 \left(1+3 B x \phi ^4\right)}{B \phi ^2 \left(1+ B x \phi ^4\right)^2}.
\end{eqnarray}
To test  the obtained  results, we would like to make contact with  the observational data including   the Planck 2018 and the recently released BICEP/Keck data\cite{r20,r21,r22}. In Fig(\ref{f1}), we illustrate the $n_s-r$ curve behaviors  for different values of  $N$, $\beta$,  and $x$.
\\
Varying  the parameters $\beta$ and $x$, these curves are plotted by implementing the Planck contour constraints. It follows from this figure that  $r$ decreases by increasing the  parameter $\beta$.
Taking large  values of $\beta$, the range of $n_s$  decreases. Moreover, the spectral index $n_s$ increases  by increasing the  parameter $x$. Fixing $\beta$ and $x$, $n_s$ increases by increasing the e-folding number $N$. It has been remarked that  the quartic  potential produces interesting  numerical values of the spectral index $n_s$ and  the tensor-to-scalar ratio $r$ for specific values of $\beta$ and $x$. In particular, this potential gives a good compatibility with the Planck results (TT,TE,EE+lowE+lensing)\cite{r20,r21,r22}. For generic values of the  involved relevant parameters $\beta$, $x$ and $N$, the range of $r$ is $[0.013, 0.123]$ while $n_s$ varies  in the range  $[0.963, 0.983]$. The quartic  potential can generate an inflation model associated with the expansion of the  universe. However, this potential does not provide an excellent compatibility with  the Planck and  the recent released BICEP/Keck data. In the next subsection, we study the small field inflation potential coupled to such a modified gravity in order  to inspect  the coupling parameter scenario impact on the compatibility with the  observational data including the recent Planck results.
 
\subsection{Small field inflation with  kinetic   coupling behaviors }
\begin{figure}[!ht]
	\centering
	\begin{tabbing}		
		\hspace{5.6cm}\= \hspace{5.6cm}\=\kill		
		\includegraphics[scale=0.60]{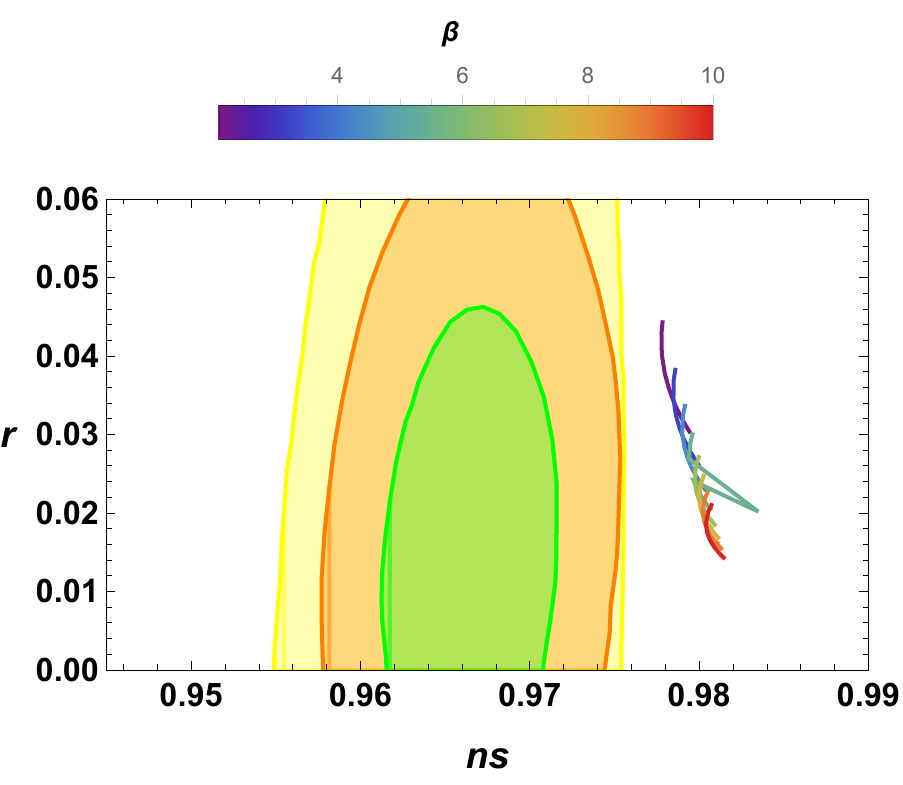} \>		
		\includegraphics[scale=0.60]{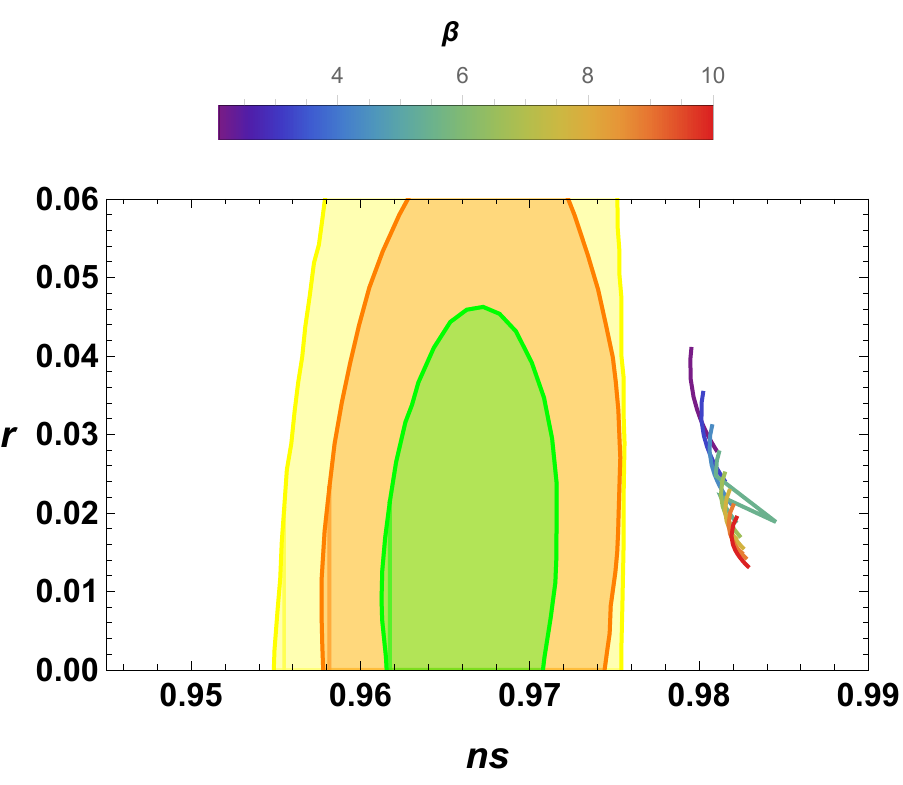} \>		
		\includegraphics[scale=0.60]{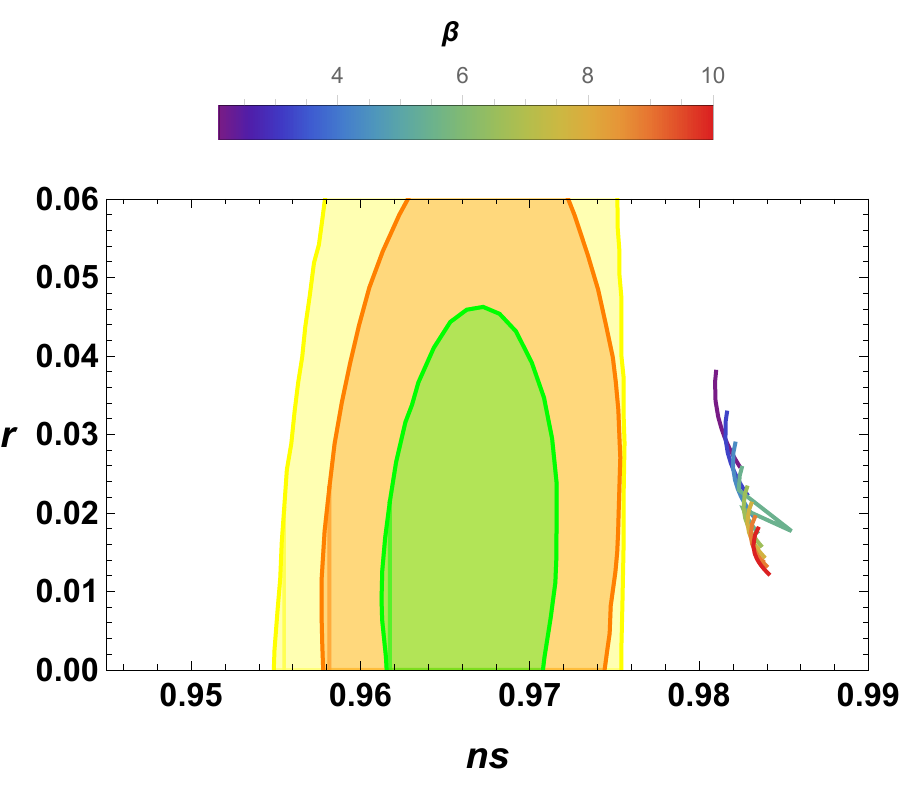} \\		
	\end{tabbing}
	\vspace{-1.5 cm}
	\caption{\it \footnotesize  The behavior of one dimensional $n_s-r$ curves  by varying the Parameter $\beta=3,\ldots, 10$, $\alpha=1$ and the parameter $y=0.001,\ldots, 1$. The plots from left to right correspond to the number e-folding  $N=60$,  $N=65$ and  $N=70$, respectively.  The green and the orange contour constraints  represent  68\% and 95\% confidentional levels of Planck results (TT,TE,EE+lowE+lensing+BK15+BAO), respectively. The yellow contour is associated with  the Planck results (TT,TE,EE+lowE+lensing).}
	\label{f2}
\end{figure}

\begin{figure}[!ht]
	
	\centering
	
	\begin{tabbing}
		
		\hspace{5.6cm}\= \hspace{5.6cm}\=\kill
		
		\includegraphics[scale=0.60]{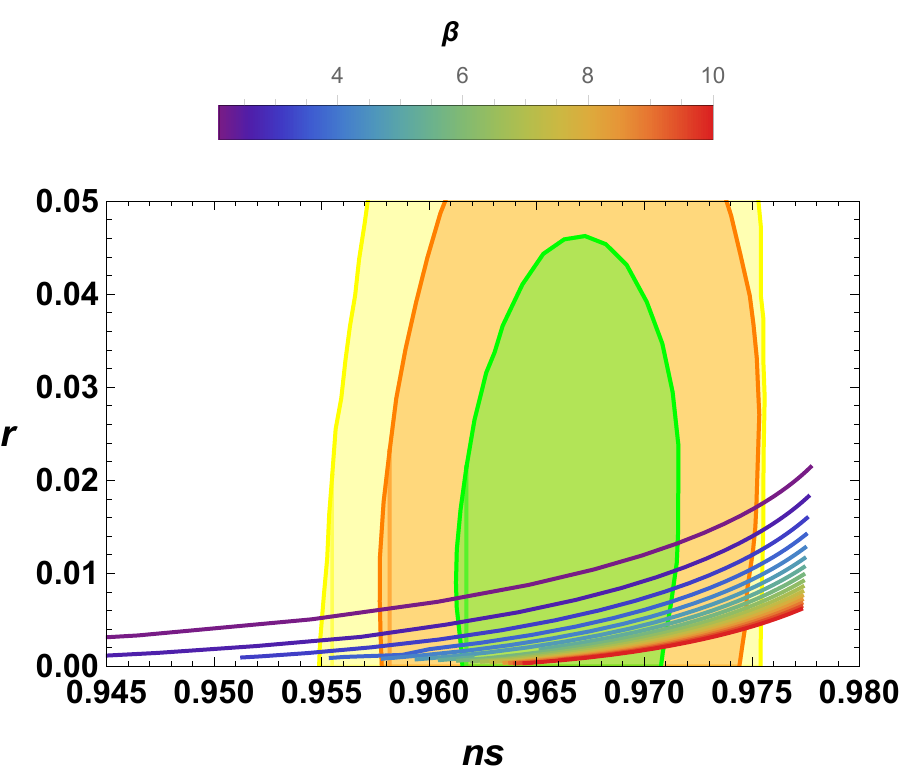} \>
		
		\includegraphics[scale=0.60]{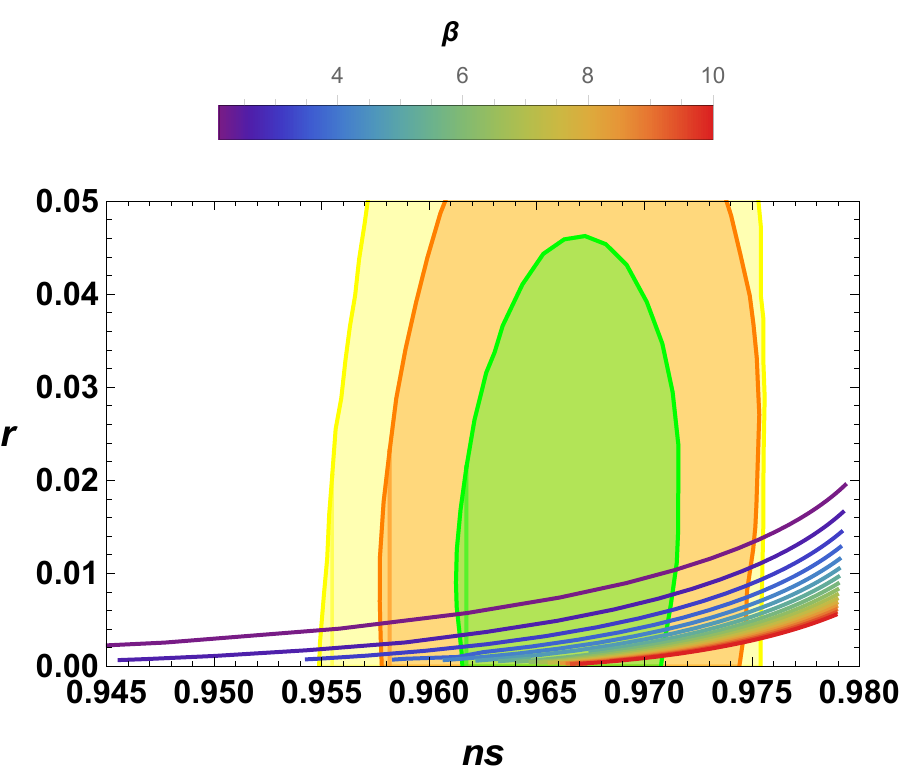} \>
		
		\includegraphics[scale=0.60]{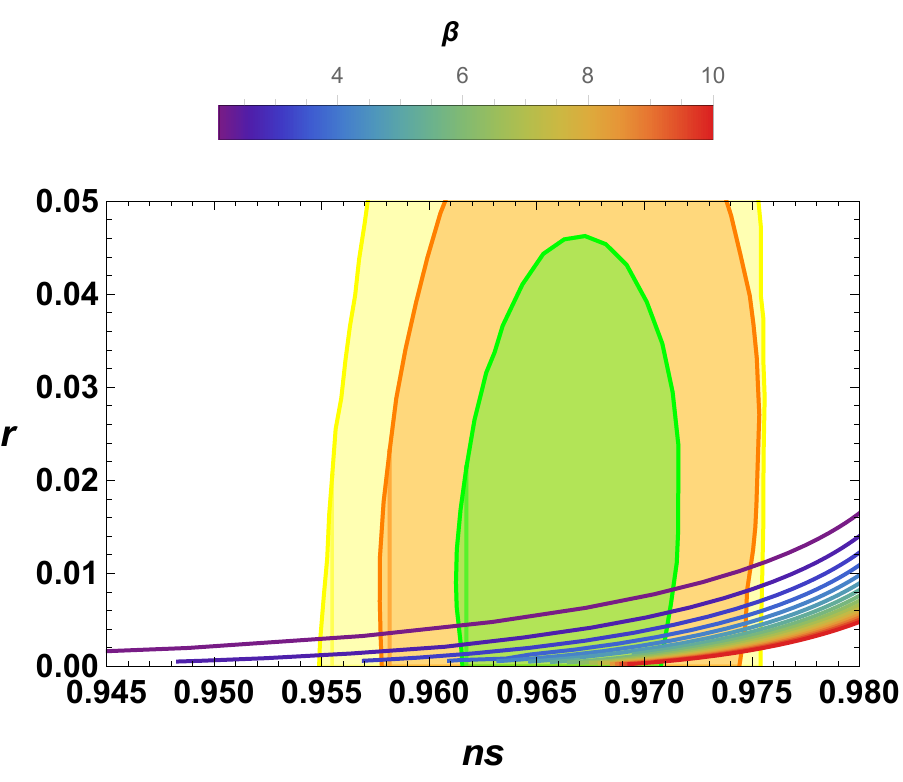} \\
		
	\end{tabbing}
	
	\vspace{-1.5 cm}
	
	\caption{\it \footnotesize  The behavior of one dimensional $n_s-r$ curves  by varying the parameter $\beta=2.1, \ldots , 10$ and the parameter $y=0.001,\ldots, 1$. The plots from left to right correspond to the number e-folding  $N=60$,  $N=65$ and  $N=70$, respectively.  The green and the orange contour constraints  represent  68\% and 95\% confidential levels of Planck results (TT,TE,EE+lowE+lensing+BK15+BAO), respectively.}
	\label{f3}
\end{figure}

Here, we consider the small field inflation  described by the following potential
\begin{equation}
	\label{V1}
	V(\phi) =V_0\left( 1- \left( \frac{\phi}{\mu}\right) ^\alpha\right) 
\end{equation} 
 depending on two parameters $\alpha$ and  $\mu$. The later is a   mass  scale parameter. However,  $\alpha$ is  a  non zero  dimensionless index \cite{r112}.  $V_0$ is a   free parameter which can be considered as $M^4$, where $M$ is a mass scale.  This  scalar potential has been used  to provide  plausible particle physics inflationary  models. A rapid examination shows that $\alpha$  and $\mu$   enlarge  now   the inflation moduli space  coordinated by $\beta, \omega, \alpha$ and $ \mu$.   Many regions and sub-spaces can be dealt  with  either by fixing certain parameters or by  identifying other ones.   Taking  $\beta=\alpha$  in the  scalar potential  can be considered as a  coupling  between the scalar field  and  the   modified  gravity through  the function $f(R,T)$.   This kind   of  couplings via    the  moduli space  could  bring  interesting   results which can be  compared  with the   Planck data. In order to compute the scalar spectral index $n_s$ and the tensor-to-scalar ratio $r$ as functions of  the involved  parameters describing the  inflation  moduli space, it is useful   to   exploit the e-folding number.    In the present work,  we have taken   a fixed value of  $\mu $  being $\mu=1$. In this way,  we get 
\begin{eqnarray}
	\label{N}
N &=&\frac{B }{2 \alpha }\left(\phi_E^2 \left(-\frac{2 \phi_E^{-\alpha } (B y+1)}{\alpha -2}+\frac{2 B y \phi_E^{\alpha }}{\alpha +2}-B y-1\right)\right) 
\\\nonumber
&-&\frac{B }{2 \alpha }\left( \phi_I^2 \left(-\frac{2\phi_I^{-\alpha } (B y+1)}{\alpha -2}+\frac{2 B y \phi_I^{\alpha }}{\alpha +2}-B y-1\right)\right)
\end{eqnarray}
where we have used $y=V_0\omega^2$.  This   e-folding number  imposes a constraint on the parameter  $\alpha$.  It  has been remarked that  this parameter should be different to $ 2$  and $-2$.
The scalar spectral index $n_s$ and  the tensor-to-scalar ratio $r$  are found to be  
\begin{eqnarray}
n_{s}  &=&1-\frac{8 \pi  \alpha ^2 \phi ^{2 \alpha -2}}{B^2 \left(1-\phi ^{\alpha }\right)^2 \left(B y \left(1-\phi ^{\alpha }\right)+1\right)}\\\nonumber	&-&\frac{\alpha  \phi ^{\alpha -2} \left(2 \left(\phi ^{\alpha }-1\right) \left(B y \left(1-\phi ^{\alpha }\right)+1\right)+\alpha  \left(2-B y \left(\phi ^{2 \alpha }+\phi ^{\alpha }-2\right)\right)\right)}{B \left(1-\phi ^{\alpha }\right)^2 \left(B y \left(1-\phi ^{\alpha }\right)-1\right)^2}
 \\
r &=&\frac{64 \pi  \alpha ^2 \phi ^{2 \alpha -2}}{B^2 \left(1-\phi ^{\alpha }\right)^2 \left(1+B\,y\,\left(1-\phi ^{\alpha }\right)\right)}.
\end{eqnarray}
 The scalar potential form and these observable quantities impose certain conditions of the parameter $\alpha$. These constraints  will be considered in the following discussions.
To inspect   the obtained  results, we should   make  contact with observational data including   the Planck 2018 and the recently released BICEP/Keck data\cite{r20,r21,r22}.\\
In Fig(\ref{f2}), we examine  the $n_s-r$ curves by varying $N$, $\beta$, $y$ and fixing $\alpha=1$.
These behaviors are plotted by implementing the Planck contour conditions  by   varying  the parameters $\beta$ and $x$ with  $\alpha=1$.    The  $n_s-r$ curves  have been examined by taking $\beta$ in the interval $ [3,10]$. It is  clear from this figure that  $r$ decreases by increasing the parameter  $\beta$. 
Taking large  values of $\beta$, the range of $n_s$  decreases. Moreover, the spectral index $n_s$ increases  by increasing the  parameter $y$. Fixing $\beta$ and $y$, $n_s$ increases by increasing $N$. For certain specific models where  the  scalar field potential and  the modified gravity $f(R,T)$  with  the kinetic couplings are not coupled,    the numerical values for the spectral index $n_s$ and  the tensor-to-scalar ratio $r$  do not belong to the   range of  the Planck data and  the recent  released BICEP/Keck data for all regions of the involved moduli space \cite{r20,r21,r22}. For generic values of the  involved  parameters including the degree polynomial of  the potential and $N$, the range of $r$ is $[0.001 , 0.043]$ while  $n_s$ varies  in the range  $[0.978,0.983]$. \\
In what follows, we couple the scalar field potential   with the   $f(R,T)$  gravity via the moduli space  by taking  $\alpha=\beta$.  This corresponds to a line in such a moduli space. 
In Fig(\ref{f3}), we illustrate the $n_s-r$ curve behaviors for different values of  $N$, $\beta$,  and  $y$.
\\
Varying  the parameters $\beta$ and $y$, these behaviors are plotted by implementing the Planck contour conditions.    The  $n_s-r$ curves  are  examined by taking $\beta$ in the interval $ [2.1,10]$. It follows from  this figure that  $r$ decreases by increasing the  gravity parameter  $\beta$.
Taking large  values of $\beta$, the range of $n_s$  decreases. Moreover, the spectral index $n_s$ increases  by increasing the  parameter $y$. Fixing $\beta$ and $y$, $n_s$ increases by increasing $N$. A close inspection  reveals  that  the coupling between  a specific scalar field potential and the modified gravity $f(R,T)$  with  the kinetic couplings provides interesting  numerical values for the spectral index $n_s$ and  the tensor-to-scalar ratio $r$ which are in a  good range of  the Planck data and  the recent  released BICEP/Keck data for all regions of the involved moduli space \cite{r20,r21,r22}. For generic values of the  relevant parameters including the  polynomial degree  of the  potential and $N$, the range of $r$ is $[0.001 , 0.014]$ while  $n_s$ varies  in the range  $[0.947,0.98]$. \\
   To end this work, we would like   to discuss  the amplitude
of  scalar power spectrum  needed  to  provide  a  viability of the proposed  theory. Following \cite{r28},  the slow-roll  analysis  can be exploited to  bring  an   expression for
  such an  amplitude. It is given approximatively  by 
  \begin{equation}
	\label{P1}
	P_{\zeta}\approx \frac{H^2}{8\pi^2\epsilon}.
\end{equation} 
To approach  such a quantity, we consider the second potential  given by Eq.(\ref{V1}) in  the  parameter   coupling scenario by taking $\alpha=\beta$. 
 The calculation gives 
 \begin{equation}
	\label{P2}
	P_{\zeta}\approx   \frac{(3 \beta +8 \pi )^3 V_0 (\frac{\phi}{\mu}) ^{2-2\beta } \left(1-(\frac{\phi}{\mu})^{\beta }\right)^3  \left(1+3 y (3 \beta +8 \pi )\left(1-(\frac{\phi}{\mu})^{\beta}\right) \right)}{96 \pi ^2 \beta ^2}.
\end{equation}
It  is observed that this  cosmological observable  depends on the matter  and  the gravity  parameters.  In some  points of  the  inflation moduli space, we could find   certain arguments with  the observational data.  Considering    $\beta =10$, $\mu=1$, $V_0=0.001$, $y=0.001$, and $\phi =0.99$,  for instance, we find  $P_{\zeta} \approx  1.96\times  10^{-6}$, being an acceptable value.

 \section{Conclusion}
In  this paper, we have  investigated the $ f(R,T)$  modified  gravity theory in  the kinetic  coupling inflation scenarios.  In particular,  we have dealt with  the form  $f(R,T)=R+2\beta T$ function by considering a positive factor for the kinetic coupling term in order to  avoid the ghost in the resulting models. Using the field variation  methods, we have obtained  the corresponding equations of motion   for a generic potential form . We have exploited  the  slow-roll  approximations  to compute  and inspect the cosmological quantities. To make contact with observational data, we have considered  two  different scalar  potentials given by  the quartic  form  $V(\phi) =\lambda \phi^4$ and  the small field inflation potential  	$V(\phi) =V_0(1- (\frac{\phi}{\mu})^\alpha)$, respectively.  Concretely, we have  dealt with  two situations described by the decoupling and   the coupling between the scalar   potential and the gravity via the  cosmological moduli space.  In the first model associated with the quartic potential, we have presented the  decoupling behavior.  In the second one corresponding to the small field potential, however, we have considered  the decoupling behavior.  Then, we  have discussed the coupling between the scalar potential and the  modified gravity by taking the line  $\alpha=\beta$ in  the moduli space.  In   such  models,  we have computed and examined   the spectral index $n_s$ and the tensor-to-scalar ratio $r$.   Precisely, we have plotted   the associated cosmological behaviors.  For three different values of  the e-folding number  $N=60,65$ and $70$,  we have found that  the  coupling  between $f(R,T)$ and the scalar  potential  via the moduli space  provides  a good agreement with observational findings.  This has been illustrated  for   the small field potential giving  an  excellent match  with the  Planck data. We anticipate that the coupling between the scalar potential and  the  modified gravity with a kinetic term  via  the function  $f(R,T)$  could provide new inflation models.    In the end of this work, we have presented   a possible  discussion on   the amplitude
of  scalar power spectrum. This  could be needed  to  provide  a  viability of the proposed  theory. Precisely,  we have considered the second potential form in the parameter coupling scenario producing acceptable values  of such a quantity  in certain points of the inflation  moduli space.

This work comes up with many open questions. A natural question is to  go beyond such  a $f(R,T)$ function by introducing  other scalar potentials in  the D-brane physics context. 
\section*{Acknowledgments}
The authors would like to thank  H. El Moumni, M. Oualaid, and M.B. Sedra  for collaborations on related subjects.  They would like also to thank the editor and the anonymous
referees for remarks,  suggestions and scientific helps.
This work is partially
supported by the ICTP through AF.

 \end{document}